\documentclass[10pt]{article}
\pdfoutput=1

\usepackage{amsmath,amsfonts,latexsym,mathtools}
\usepackage{graphicx}
\usepackage[hidelinks]{hyperref} 
\usepackage[text={16cm,22cm},top=3cm]{geometry}  
\numberwithin{equation}{section}

\usepackage[usenames,dvipsnames,svgnames]{xcolor}

\hypersetup{
    colorlinks,
    linkcolor={blue!50!black},
    citecolor={blue!50!black},
    urlcolor={blue!50!black}
}

\usepackage[all]{hypcap}  

\usepackage[super]{nth}

\usepackage{comment}

\usepackage{tabularx}
\usepackage{hhline}
\newcolumntype{Y}{>{\centering\arraybackslash}X}

\allowdisplaybreaks[1]

\newcommand{\nc}{\newcommand}
\nc{\numberthis}{\addtocounter{equation}{1}\tag{\theequation}}

\nc{\dsp}[1]{^\mathrm{#1}}
\nc{\lb}{\left (}
\nc{\rb}{\right )}
\nc{\lset}{\left \{}
\nc{\rset}{\right \}}
\nc{\eqtext}[1]{\quad \text{#1} \quad}
\nc{\lsq}{\left [}
\nc{\rsq}{\right ]}

\nc{\half}{\frac{1}{2}}
\nc{\RA}{\quad \Rightarrow \quad}
\nc{\sech}[1]{\mathrm{sech} \lb #1 \rb}
\nc{\coshn}[2]{\mathrm{cosh}^{#1} \lb #2 \rb}
\nc{\sechn}[2]{\mathrm{sech}^{#1} \lb #2 \rb}
\nc{\tanhn}[2]{\mathrm{tanh}^{#1} \lb #2 \rb}
\nc{\arccosh}[1]{\mathrm{arccosh} \lb #1 \rb}
\nc{\pqsum}{P \oplus Q}
\nc{\ob}[1]{\overbrace{#1}}
\nc{\ub}[1]{\underbrace{#1}}
\nc{\field}[1]{\mathbb{#1}}
\nc{\inflim}{_{-\infty}^{\infty}}

\nc{\dd}[1]{\; \mathrm{d} #1}
\nc{\diff}[2]{\frac{\mathrm{d} #1}{\mathrm{d} #2}}
\nc{\diffn}[3]{\dfrac{\mathrm{d}^{#1} #2}{\mathrm{d} #3^{#1}}}
\nc{\pdd}[1]{\; \partial #1}
\nc{\pdiff}[2]{\frac{\partial #1}{\partial #2}}
\nc{\pdiffn}[3]{\dfrac{\partial^{#1} #2}{\partial #3^{#1}}}

\nc{\inteps}[2]{\int_{-\epsilon}^{\epsilon} #1 \dd{#2}}
\nc{\limeps}{\lim_{\epsilon \rightarrow 0^{+}} }

\nc{\nt}{\newtheorem}
\nc{\ntc}{\newtheorem*}

\nt{theorem}{Theorem}[section]
\nt{definition}{Definition}[section]
\nt{exa}{Example}[section]
\nt{lem}{Lemma}[section]
\nt{cor}{Corollary}[section]
\nt{pro}{Proposition}[section]

\begin{document}

\renewcommand*{\thefootnote}{\fnsymbol{footnote}}

\begin{center}
{\LARGE Modelling of nonlinear wave scattering \\ in a delaminated elastic bar\footnote{Accepted for publication in \textbf{Proc. Roy. Soc. A (2015): http://dx.doi.org/10.1098/rspa.20150584 }}} \\[2.5em]
{\large K. R. Khusnutdinova\footnote{Corresponding author. Tel: +44 (0)1509 228202. Fax: +44 (0)1509 223969.}, M. R. Tranter} \\[1.5em]
{Department of Mathematical Sciences, Loughborough University,\\ Loughborough, LE11 3TU, United Kingdom} \\[1.5em]
{K.Khusnutdinova@lboro.ac.uk} \\
{M.Tranter@lboro.ac.uk}
\end{center}

\renewcommand*{\thefootnote}{\arabic{footnote}}
\setcounter{footnote}{0}

\begin{abstract}
Integrity of layered structures, extensively used in modern industry, strongly depends on the quality of their interfaces; poor adhesion or delamination can lead to a failure of the structure. Can nonlinear waves help us to control the quality of layered structures?

In this paper we numerically model the dynamics of a long longitudinal strain solitary wave in a split, symmetric layered bar.  The recently developed analytical approach, based on matching two asymptotic multiple-scales expansions and the integrability theory of the KdV equation by the Inverse Scattering Transform, is used to develop an effective semi-analytical numerical approach for these types of problems.
We also employ a direct finite-difference method and compare the numerical results with each other, and with the analytical predictions.  The numerical modelling confirms that delamination causes fission of an incident solitary wave and, thus, can be used to detect the defect. 
\end{abstract}

\section{Introduction}
The birth of the concept of a {\it soliton} is ultimately linked with the Boussinesq equation appearing in the continuum approximation of the famous Fermi-Pasta-Ulam problem \cite{Zabusky65}. Although the vast majority of the following studies of solitons were devoted to waves in fluids and nonlinear optics (see, for example, \cite{Johnson97,Yang10, Ablowitz11} and references therein) it was also  shown, within the framework of the nonlinear dynamic elasticity theory, that Boussinesq-type equations can be used to model the propagation of long nonlinear longitudinal bulk strain waves in rods and plates (see   \cite{Samsonov01, Porubov03}). The existence of longitudinal bulk strain solitons in these solid waveguides, predicted by the model equations, was confirmed by experiments \cite{Dreiden88, Samsonov98, Semenova05}.  Very recently the theoretical and experimental studies were extended to some types of adhesively bonded layered bars \cite{Khusnutdinova08,Khusnutdinova09,Dreiden08,Dreiden11,Dreiden14a} and  thin-walled cylindrical shells \cite{Dreiden14}. 

Unusually slow decay of the observed solitons in some polymeric waveguides makes them an attractive candidate for the introscopy of the waveguides, in addition to the existing methods.
Recently, we analytically studied scattering of the longitudinal bulk strain soliton by the inhomogeneity modelling delamination in a symmetric layered bar with perfect bonding \cite{Khusnutdinova08}. The developed theory was supported by experiments \cite{Dreiden10,Dreiden12}. In order to extend the research to other, more complicated types of layered structures, the analytical and experimental studies need to be complemented with numerical modelling. The direct numerical modelling of even the simplest problem proved to be rather expensive. The development of numerical schemes based on analytical results is an actively developing direction of research (e.g. asymptotic techniques have been used to improve the efficiency of numerics for tsunamis in \cite{Berry05}, the Inverse Scattering Transform (IST) has been used for the KdV in \cite{Trogdon12}, Pad\'{e} approximants have been used to reduce the Gibbs phenomenon in \cite{Andrianov14}, series expansions in the vertical excursions of the interface and bottom topography are used in the three-dimensional modelling of tidally driven internal wave formation in \cite{Grue15}). 

The aim of this paper is to develop an efficient semi-analytical numerical approach based on the weakly nonlinear analysis of the problem, which could be used to model the scattering of nonlinear waves on the defects. The developed approach can be applied to other problems of this type. For example,  a system of coupled Boussinesq-type equations was derived to describe longitudinal waves in a layered waveguide with soft bonding \cite{Khusnutdinova09}.  In \cite{Khusnutdinova11}  we constructed a hierarchy of weakly nonlinear solutions of the initial-value problem for localised initial data  in terms of solutions of Ostrovsky-type equations \cite{Ostrovsky78} for unidirectional waves, which opens the way for the application of the method developed in the present paper to the study of the scattering of nonlinear waves in other delaminated layered structures.

The problem of calculating the reflected and transmitted waves from a wave incident on an interface between two media has been discussed in several areas. In fluids, it is known that when a surface soliton passes through an area of rapid depth variation, an incident solitary wave evolves into a group of solitons \cite{Pelinovsky71, Tappert71, Johnson73}. There is also a similar effect for internal waves \cite{Djordjevic78, Grimshaw08, Maderich10}. The problem of collimated light beams incident on an interface separating two linear dielectric media has been extensively studied \cite{Lotsch68, Horowitz71, Ra73}. This can be extended to nonlinear dielectric media and theoretical results have been found \cite{Aceves89}. Interface Sturm-Liouville systems have been studied in \cite{Barrett88} and applications of these systems have been discussed in \cite{Sangren60, Wylie82}. It is worth noting that the systems considered in these examples often have simple boundary conditions, while our problem takes account of continuity of stress.

The paper is organised as follows. In  Section \ref{sec:WNL}, we briefly overview the problem formulation for the scattering of an incident soliton in a symmetric layered bar with delamination, and the weakly nonlinear solution  of the problem \cite{Khusnutdinova08}. The approach gives rise to three KdV equations describing the behaviour of the leading order incident, transmitted and reflected waves, and terms for the higher order corrections. We describe fission of the transmitted and reflected strain solitary waves and establish predictions for the number of solitary waves present in each section of the bar. In Section \ref{sec:Num}, we present two numerical schemes. Section \ref{sec:FDMNum} describes the finite-difference approximations for the problem, with a ``kink'' as the initial condition for the displacement (corresponding to the strain soliton). The system is treated as two boundary value problems and the finite-difference approximation of the system produces two tridiagonal matrices with a nonlinear condition at $x=0$ (the common point in both boundary value problems). 
In Section \ref{sec:KNum} we introduce a semi-analytical numerical scheme, which is based upon the weakly nonlinear solution discussed in Section \ref{sec:WNL} . The three KdV equations are solved numerically using a Strong Stability Preserving Runge-Kutta (SSPRK) scheme proposed in \cite{Gottlieb09}. In Section \ref{sec:Res} we compare the results of the numerical schemes from Section \ref{sec:Num}, subsections \ref{sec:FDMNum} and \ref{sec:KNum}, against each other and the predicted results, and the agreement is checked for various configurations of the delaminated bar. Finally, we conclude our findings in Section \ref{sec:Conc} and discuss potential extensions of this work, pertaining to other types of bonding of layered structures, described by coupled regularised Boussinesq (crB) equations \cite{Khusnutdinova09}.

\section{Weakly Nonlinear Solution}
\label{sec:WNL}
\begin{figure}
\begin{center}
\includegraphics[width=0.6\linewidth, trim = 0mm 10mm 0mm 10mm]{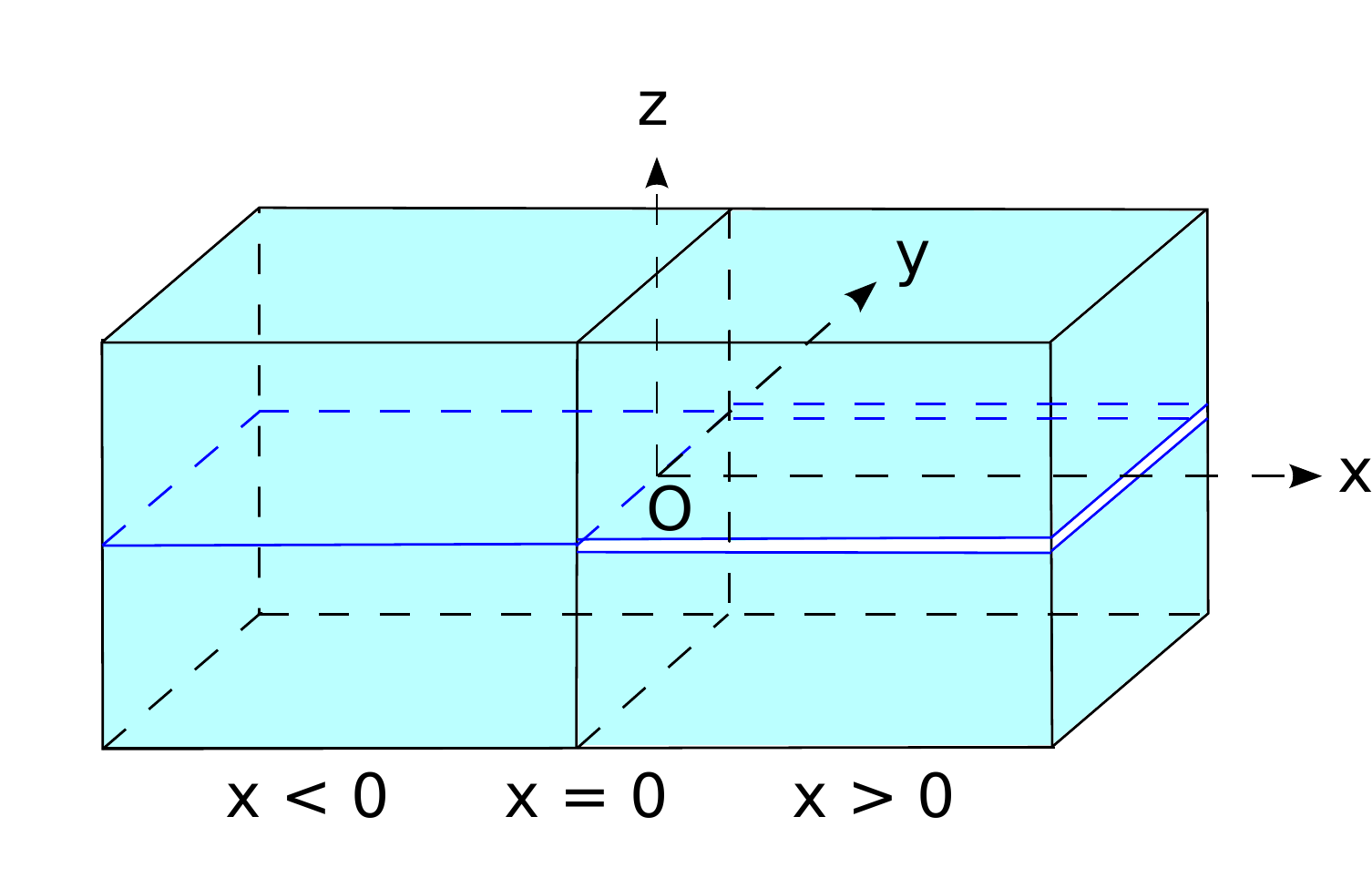}
\caption{Two-layered symmetric bar with delamination at $x>0$. }
\label{DelamBar}
\end{center}
\end{figure}

We consider the scattering of a long longitudinal strain solitary wave in a perfectly bonded two-layered bar with delamination at $x>0$ (see  Figure \ref{DelamBar}). The material of the layers is assumed to be the same (symmetric bar), while the material to the left and to the right of the $x=0$ cross-section can be different. The problem is described by the following set of regularised nondimensional equations \cite{Khusnutdinova08}
\begin{align}
u_{tt}^{-} - u_{xx}^{-} &= \epsilon \lsq -12 u_x^{-} u_{xx}^{-} + 2u_{ttxx}^{-} \rsq, \quad x < 0, \notag \\
u_{tt}^{+} - c^2 u_{xx}^{+} &= \epsilon \lsq -12 \alpha u_x^{+} u_{xx}^{+} + 2\frac{\beta}{c^2} u_{ttxx}^{+} \rsq, \quad x > 0, \label{SB}
\end{align}
with appropriate initial conditions
\begin{equation}
u^{\pm}(x,0) = F^{\pm}(x),
\label{SBIC}
\end{equation}
and associated continuity conditions
\begin{align}
u^{-}|_{x=0} &= u^{+}|_{x=0}, \label{Con1} \\
\left. u_x^{-} + \epsilon \lsq -6 \lb u_x^{-} \rb^2  + 2u_{ttx}^{-} \rsq \right |_{x=0} &= \left. c^2 u_{x}^{+} + \epsilon \lsq -6\alpha \lb u_{x}^{+} \rb^2 + 2\frac{\beta}{c^2} u_{ttx}^{+} \rsq \right |_{x=0}, \label{Con2}
\end{align}
where $c, \alpha, \beta$ are constants defined by the geometrical and physical parameters of the structure, while $\epsilon$ is the small wave amplitude parameter. The functions $u^{-}(x,t)$  and $u^{+}(x,t)$ describe displacements in the bonded and delaminated areas of the structure respectively. Condition \eqref{Con1} is continuity of longitudinal displacement, while condition \eqref{Con2} is the continuity of stress.
 In order to reduce the number of parameters in our numerical experiments, in what follows we will use the values presented in \cite{Khusnutdinova08}, namely that $\alpha = 1$ and
\begin{equation}
\beta = \frac{n^2 + k^2}{n^2 \lb 1 + k^2 \rb},
\label{betaval}
\end{equation}
where $n$ represents the number of layers in the structure and $k$ is defined by the geometry of the waveguide. Referring to Figure \ref{DelamBar}, the cross-section $x=0$ has width $2a$ and the height of each layer is $2b/n$.  In terms of these values, $k = b/a$ and, as there are two layers in this example, $n=2$. In Section \ref{sec:Res} we will consider various configurations of the bar.

\subsection{Derivation of KdV Equations}
Let us consider a weakly nonlinear solution to \eqref{SB} of the form \cite{Khusnutdinova08}
\begin{align*}
u^{-} &= I \lb \xi_{-}, X \rb + R \lb \eta_{-}, X \rb + \epsilon P \lb \xi_{-}, \eta_{-}, X \rb + O \lb \epsilon^2 \rb, \\
u^{+} &= T \lb \xi_{+}, X \rb + \epsilon Q \lb \xi_{+}, \eta_{+}, X \rb + O \lb \epsilon^2 \rb,
\end{align*}
where the characteristic variables are given by
\begin{align*}
\xi_{-} &= x - t, &\xi_{+} &= x - ct, &\eta_{-} &= x + t, &\eta_{+} &= x + ct, &X &= \epsilon x.
\end{align*}
Here, the functions $I \lb \xi_{-}, X \rb$, $R \lb \eta_{-}, X \rb$, $T \lb \xi_{+}, X \rb$ describe the leading order incident, reflected and transmitted waves respectively, while the functions $P \lb \xi_{-}, \eta_{-}, X \rb$ and $Q \lb \xi_{+}, \eta_{+}, X \rb$ describe the higher-order corrections. Substituting this weakly nonlinear solution into \eqref{SB} the system is satisfied at leading order, while at $O \lb \epsilon \rb$ we have
\begin{align}
-2P_{\xi_{-} \eta_{-}} =& \lb I_{X} - 3I_{\xi_{-}}^2 + I_{\xi_{-} \xi_{-} \xi_{-}} \rb_{\xi_{-}} + \lb R_{X} - 3R_{\eta_{-}}^2 + R_{\eta_{-} \eta_{-} \eta_{-}} \rb_{\eta_{-}} \notag \\
& - 6 \lb R I_{\xi_{-}} + I R_{\eta_{-}} \rb_{\xi_{-} \eta_{-}}. \label{Pexp}
\end{align}
To leading order the right-propagating incident wave
\begin{equation}
I = \int \tilde{I} \dd{\xi_{-}},
\end{equation}
is defined by the solution of the KdV equation
\begin{equation}
\tilde{I}_{X} - 6 \tilde{I} \tilde{I}_{\xi_{-}} + \tilde{I}_{\xi_{-} \xi_{-} \xi_{-}} = 0.
\label{IKdV}
\end{equation}
Similarly the reflected wave
\begin{equation}
R = \int \tilde{R} \dd{\eta_{-}},
\end{equation}
satisfies the KdV equation
\begin{equation}
\tilde{R}_{X} - 6 \tilde{R} \tilde{R}_{\eta_{-}} + \tilde{R}_{\eta_{-} \eta_{-} \eta_{-}} = 0.
\label{RKdV}
\end{equation}
Substituting these conditions into \eqref{Pexp} and integrating with respect to both characteristic variables, we obtain an expression for higher-order terms satisfying
\begin{equation}
P = 3 \lb R I_{\xi_{-}} + I R_{\eta_{-}} \rb + \phi \lb \xi_{-}, X \rb + \psi \lb \eta_{-}, X \rb,
\label{Peq}
\end{equation}
where $\phi \lb \xi_{-}, X \rb$, $\psi \lb \eta_{-}, X \rb$ are arbitrary functions. The first radiation condition requires that the solution to the problem should not change the incident wave (see \cite{Khusnutdinova08} for the discussion of the radiation conditions). In our problem the incident wave is  the solitary wave solution of the KdV equation \eqref{IKdV}, with corrections at $O \lb \epsilon^2 \rb$.
Therefore, the radiation condition implies that $\phi \lb \xi_{-}, X \rb = 0$. We find $\psi \lb \eta_{-}, X \rb$ from conditions \eqref{Con1} and \eqref{Con2} later. A similar calculation for the second equation in \eqref{SB} gives
\begin{equation}
-2c^2 Q_{\xi_{+} \eta_{+}} = \lb c^2 T_{X} - 3 T_{\xi_{+}}^2 + \beta T_{\xi_{+} \xi_{+} \xi_{+}} \rb_{\xi_{+}}.
\end{equation}
We are looking for the leading order transmitted wave satisfying
\begin{equation}
T = \int \tilde{T} \dd{\xi_{+}},
\end{equation}
where
\begin{equation}
\tilde{T}_{X} - \frac{6}{c^2} \tilde{T} \tilde{T}_{\xi_{+}} + \frac{\beta}{c^2} \tilde{T}_{\xi_{+} \xi_{+} \xi_{+}} = 0,
\label{TKdV}
\end{equation}
and higher order corrections are given by
\begin{equation}
Q = q \lb \xi_{+}, X \rb + r \lb \eta_{+}, X \rb,
\label{Qeq}
\end{equation}
where $q \lb \xi_{+}, X \rb$, $r \lb \eta_{+}, X \rb$ are arbitrary functions. The second radiation condition states that if the incident wave is right-propagating, then there should be no left-propagating waves in the transmitted wave field. Thus, $r \lb \eta_{+}, X \rb = 0$. From \eqref{Con1} we see
\begin{equation*}
u_{t}^{-}|_{x=0} = u_{t}^{+}|_{x=0},
\end{equation*}
so to leading order we have
\begin{equation}
I_{\xi_{-}} |_{x=0} - R_{\eta_{-}} |_{x=0} = c T_{\xi_{+}} |_{x=0},
\label{ICon1}
\end{equation}
and at $O \lb \epsilon \rb$
\begin{equation}
\psi_{\eta_{-}} |_{x=0} + c q_{\xi_{+}} |_{x=0} = 3 \lb I_{\xi_{-} \xi_{-}} R - I R_{\eta_{-} \eta_{-}} \rb |_{x=0} = f \lb t, X \rb |_{x=0}.
\label{IDCon1}
\end{equation}
Similarly for \eqref{Con2} we have, to leading order
\begin{equation}
I_{\xi_{-}} |_{x=0} + R_{\eta_{-}} |_{x=0} = c^2 T_{\xi_{+}} |_{x=0},
\label{ICon2}
\end{equation}
and at $O \lb \epsilon \rb$
\begin{align}
\psi_{\eta_{-}} |_{x=0} - c^2 q_{\xi_{+}} |_{x=0} &= \lsq - \lb I_{X} - 6 I_{\xi_{-}}^{2} + 2 I_{\xi_{-} \xi_{-} \xi_{-}} \rb - \lb R_{X} - 6 R_{\eta_{-}}^{2} + 2 R_{\eta_{-} \eta_{-} \eta_{-}} \rb \right. \notag \\
& \left. \quad +~ c^2 T_{X} - 6 T_{\xi_{+}}^{2} + 2 \beta T_{\xi_{+} \xi_{+} \xi_{+}} - 3 \lb I R_{\eta_{-} \eta_{-}} + I_{\xi_{-} \xi_{-}} R \rb \rsq |_{x=0} \notag \\
&= g \lb t, X \rb |_{x=0}. \label{IDCon2}
\end{align}
Using \eqref{ICon1} and \eqref{ICon2} we can derive initial conditions for the previously derived KdV equations, defining both reflected and transmitted ``strain'' waves at $x=0$ in terms of the incident wave i.e.
\begin{equation}
\tilde{R} |_{x=0} = C_{R} \tilde{I} |_{x=0}, \quad \tilde{T} |_{x=0} = C_{T} \tilde{I} |_{x=0},
\label{IIC}
\end{equation}
where
\begin{equation}
C_{R} = \frac{c - 1}{c + 1},
\label{CR}
\end{equation}
and
\begin{equation}
C_{T} = \frac{2}{c \lb c + 1 \rb},
\label{CT}
\end{equation}
are the reflection and transmission coefficients respectively. It is worth noting that, if the materials are the same in the bonded and delaminated areas $\lb c =  1 \rb$, then $C_{T} = 1$, $C_{R} = 0$ and there is no leading order reflected wave.

\noindent We can now simplify \eqref{IDCon1} and \eqref{IDCon2} using the KdV equations \eqref{IKdV}, \eqref{RKdV}, \eqref{TKdV} and relations \eqref{IIC} to obtain
 \begin{align}
f \lb t, X \rb |_{x=0} &= \lsq 3 \lb R + C_{R} I \rb I_{\xi_{-} \xi_{-}} \rsq |_{x=0}, \label{feq} \\
g \lb t, X \rb |_{x=0} &= \lsq 3 \lb 1 + C_{R}^{2} - C_{T}^{2} \rb I_{\xi_{-}}^{2} - \lb 1 + C_{R} -   \beta c^{-2} C_{T} \rb I_{\xi_{-} \xi_{-} \xi_{-}} \right. \notag \\
& \left. \qquad - 3 \lb R - C_{R} I \rb I_{\xi_{-} \xi_{-}} \rsq |_{x=0}. \label{geq}
\end{align}
Considering \eqref{IDCon1} and \eqref{IDCon2} we have
\begin{align}
\psi_{\eta_{-}} |_{x=0} &= \frac{1}{1 + c} \lsq cf \lb t, X \rb + g \lb t, X \rb \rsq |_{x=0}, \label{psieq} \\
q_{\xi_{+}} |_{x=0} &= \frac{1}{c \lb 1 + c \rb} \lsq f \lb t, X \rb - g \lb t, X \rb \rsq |_{x=0}. \label{qeq}
\end{align}
The functions $f, g$ are now completely defined in terms of the leading order incident, reflected and transmitted waves. We restore the dependence of $f$ and $g$ on their respective characteristic variables to obtain
\begin{align}
\psi \lb \eta_{-}, X \rb &= \frac{1}{1 + c} \int \lsq cf \lb \eta_{-}, X \rb + g \lb \eta_{-}, X \rb \rsq \dd{\eta_{-}}, \label{psieq2} \\
q \lb \xi_{+}, X \rb &= \frac{1}{c \lb 1 + c \rb} \int \lsq f \lb -\frac{\xi_{+}}{c}, X \rb - g \lb -\frac{\xi_{+}}{c}, X \rb \rsq \dd{\xi_{+}}. \label{qeq2}
\end{align}
The constants of integration in the KdV equations and \eqref{psieq2}, \eqref{qeq2}, should be found from additional physical conditions.

\subsection{Fission}
\label{sec:Fis}
We can rewrite the transmitted wave equation in the canonical form
\begin{equation*}
U_{\tau} - 6 U U_{\chi} + U_{\chi \chi \chi} = 0,
\end{equation*}
via the change of variables
\begin{equation}
U = \frac{1}{\beta} \tilde{T}, \quad \tau = \frac{\beta}{c^2} X, \quad \chi = \xi_{+}.
\end{equation}
The method of the Inverse Scattering Transform \cite{GGKM67} can be used to determine the solution for the KdV equation. We make a similar approximation to what was done in \cite{Johnson73}, for a soliton moving into a region with different properties, neglecting some short waves as the soliton moves over the $x=0$ cross-section. We define the transmitted wave field by the spectrum of the Schr\"{o}dinger equation
\begin{equation}
\Psi_{\chi \chi} + \lsq \lambda - U(\chi) \rsq \Psi = 0,
\label{Schro}
\end{equation}
where the potential is given by
\begin{equation}
U(\chi) = -A ~\sechn{2}{\frac{\chi}{l}}, \quad A = \frac{v}{\beta c \lb 1 + c \rb}, \quad l = \frac{2c}{\sqrt{v}}.
\label{UPot}
\end{equation}
The sign of $A$ will determine if any solitary waves are present in the transmitted wave field. If $A<0$, the transmitted wave field will not contain any solitons and the initial pulse will degenerate into a dispersive wave train. However when $A>0$, there will always be at least one discrete eigenvalue, corresponding to at least one solitary wave in the transmitted wave field, and accompanying radiation determined by the continuous spectrum. It is worth noting that, in some cases, more than one secondary soliton can be produced from the single initial soliton, referred to as fission of a soliton \cite{Pelinovsky71, Tappert71, Johnson73}.

\noindent The discrete eigenvalues of \eqref{Schro} take the form (see, for example, \cite{Landau13})
\begin{equation}
\lambda = -k_{n}^2, \eqtext{where} k_{n} = \frac{1}{2l} \lsq \lb 1 + 4Al^2 \rb^{1/2} - \lb 2n-1 \rb \rsq, \quad n = 1, 2, \dots , N.
\label{Eig}
\end{equation}
The number of solitary waves produced in the delaminated area, $N$, is given by the largest integer satisfying the inequality
\begin{equation}
N < \frac{1}{2} \lb \sqrt{1 + \frac{4 \delta^2}{\pi^2}} + 1 \rb,
\label{del}
\end{equation}
where
\begin{equation*}
\delta = \pi \sqrt{Al} = 2 \pi \sqrt{\frac{c}{\beta \lb 1 + c \rb}}.
\end{equation*}
In the above, the parameters $ \beta$ and $c$ depend on the properties of the material and the geometry of the waveguide. We can see from \eqref{del} that, for small $\delta$, there will always be one soliton while, as $\delta$ increases, more solitons will emerge. As $\tau \rightarrow +\infty$, the solution will evolve into a train of solitary waves, ordered by their heights, propagating to the right and some dispersive radiation (a dispersive wave train) propagating to the left (in the moving reference frame) i.e.
\begin{equation}
U \sim - \sum_{n=1}^{N} 2 k_{n}^2 ~\sechn{2}{k_{n} \lb \chi - 4 k_{n}^2 \tau - \chi_{n} \rb} + ~\text{radiation},
\end{equation}
where $\chi_{n}$ is the phase shift.

Is soliton fission possible if the waveguide is made of one and the same material? This requires setting $c=1$. We find that fission can occur and is dependent upon the geometry of the waveguide. Recall the expression for $\beta$ given in \eqref{betaval}, where $n$ is the number of layers and $k=b/a$.
The number of solitary waves produced is the largest integer satisfying
\begin{equation*}
N <  \frac{1}{2} \lb \sqrt{1 + 8n^2 \frac{1 + k^2}{n^2+k^2}} +1 \rb .
\end{equation*}
This gives rise to a series of predictions based upon the value of $\beta$ and these will be checked in Section \ref{sec:Res}.

A similar description can be made for the reflected wave, which is already written in canonical form in \eqref{RKdV}, making use of the ``initial condition'' and reflection coefficient as presented in \eqref{IIC} and \eqref{CR} respectively. The wave field is defined by the spectrum of the Schr\"{o}dinger equation \eqref{Schro}, where the potential $U$ is given by
\begin{equation*}
U \lb \chi \rb = - B ~\sechn{2}{\frac{\chi}{m}}, \quad B = \frac{v}{2} C_R = \frac{v \lb c - 1 \rb}{2 \lb c + 1 \rb}, \quad m = \frac{2}{\sqrt{v}},
\end{equation*}
where we have used \eqref{CR}. It is clear that the sign of $B$ is dependent upon the sign of the reflection coefficient $C_R$. If $c < 1$, then $B$ is negative and the reflected wave field does not contain any solitary waves. The initial pulse will degenerate into a dispersive wave train. For $c > 1$, $B$ is positive and there will be at least one solitary wave present in the reflected wave field, accompanied by radiation. The solitary waves can be described using formulae \eqref{Eig} and \eqref{del} and making the change $A \rightarrow B$, and $l \rightarrow m$.

If $c = 1$ (the structure is of one and the same material) then $C_R = 0$ and there is no leading order reflected wave.

\section{Numerical Schemes}
\label{sec:Num}
To numerically solve equations \eqref{SB}, with continuity conditions \eqref{Con1} and \eqref{Con2}, we consider two numerical schemes. The first builds on the scheme for a single Boussinesq equation \cite{Khusnutdinova11}, treating the problem as two boundary value problems (BVPs) with the continuity conditions defining the interaction between the two problems. The second approach is a semi-analytical method that solves the equations derived in Section \ref{sec:WNL} using a Strong Stability Preserving Runge-Kutta (SSPRK) scheme (see \cite{Spiteri03, Ruuth06, Gottlieb09} for method and \cite{Obregon14} for examples).

\subsection{Direct Finite-Difference Scheme}
\label{sec:FDMNum}
Firstly we consider the Boussinesq-type equations posed in Section \ref{sec:WNL}. Discretising the domain $\lsq -L, L \rsq \times \lsq 0, T \rsq$ into a grid with equal spacings $h = \Delta x$ and $\kappa = \Delta t$, the analytical solution $u^{\pm} \lb x, t \rb$ is approximated by the exact solution of the finite difference scheme $u^{\pm} \lb ih, j \kappa \rb$, denoted $u_{i,j}^{\pm}$. We make use of first order and second order central difference approximations in the main equations, namely
\begin{equation*}
u_x^{\pm} = \frac{u_{i+1,j}^{\pm} - u_{ i-1,j}^{\pm}}{2h}, \quad u_{xx}^{\pm} = \frac{u_{i+1,j}^{\pm} - 2u_{i,j}^{\pm} + u_{i-1,j}^{\pm}}{h^2}, \quad u_{tt}^{\pm} = \frac{u_{i,j+1}^{\pm} - 2u_{i,j}^{\pm} + u_{i,j-1}^{\pm}}{\kappa^2},
\end{equation*}
and the approximation for $u_{ttxx}^{\pm}$ can be obtained iteratively using the approximations for $u_{tt}^{\pm}$ and $u_{xx}^{\pm}$. To simplify the obtained expressions, we introduce the notation
\begin{equation*}
D_{xx} \lb u_{i,j}^{-} \rb = u_{i+1,j}^{-} - 2u_{i,j}^{-} + u_{i-1,j}^{-}.
\end{equation*}
Substituting these into system \eqref{SB} gives
\begin{align}
&-2 \epsilon u_{i+1,j+1}^{-} + \lb 4 \epsilon + h^2 \rb u_{i,j+1}^{-} - 2 \epsilon u_{i-1,j+1}^{-} = \lb \kappa^2 - 4\epsilon \rb D_{xx} \lb u_{i,j}^{-} \rb \notag \\
&+ 2h^2 u_{i,j}^{-} - \frac{6 \epsilon \kappa^2}{h} \lsq \lb u_{i+1,j}^{-} \rb^2 - \lb u_{i-1,j}^{-} \rb^2 - 2u_{i+1,j}^{-} u_{i,j}^{-} + 2u_{i,j}^{-} u_{i-1,j}^{-} \rsq \notag \\
&+ 2 \epsilon u_{i+1,j-1}^{-} - \lb 4\epsilon + h^2 \rb u_{i,j-1}^{-} + 2 \epsilon u_{i-1,j-1}^{-},
\label{fd_uminus}
\end{align}
and
\begin{align}
&- \frac{2\epsilon \beta}{c^2} u_{i+1,j+1}^{+} + \lb \frac{4\epsilon\beta}{c^2} + h^2 \rb u_{i,j+1}^{+} - \frac{2\epsilon\beta}{c^2} u_{i-1,j+1}^{+} = \lb \kappa^2 c^2 -  \frac{4\epsilon\beta}{c^2} \rb D_{xx} \lb u_{i,j}^{+} \rb \notag \\
& + 2h^2 u_{i,j}^{+} -
\frac{6\epsilon \kappa^2}{h} \lsq \lb u_{i+1,j}^{+} \rb^2 - \lb u_{i-1,j}^{+} \rb^2 - 2 u_{i+1,j}^{+} u_{i,j}^{+} + 2 u_{i,j}^{+} u_{i-1,j}^{+} \rsq \notag \\
&+  \frac{2\epsilon\beta}{c^2} u_{i+1,j-1}^{+} - \lb  \frac{4\epsilon\beta}{c^2} + h^2 \rb u_{i,j-1}^{+} +  \frac{2\epsilon\beta}{c^2} u_{i-1,j-1}^{+}. \label{fd_uplus}
\end{align}
Assuming the domain can be discretised, we denote
\begin{equation}
N = \frac{L}{h},
\end{equation}
and therefore continuity condition \eqref{Con1} becomes
\begin{equation}
u_{N,j+1}^{-} = u_{N,j+1}^{+}.
\label{fd_Con1}
\end{equation}
In the continuity condition \eqref{Con2}, we make use of the central difference approximations presented above, and introduce ``ghost points'' of the form $u_{N+1,j+1}^{-}$ and $u_{N-1,j+1}^{+}$. Therefore, \eqref{Con2} becomes
\begin{align}
&4 h \kappa^2 \lb u_{N+1,j+1}^{-} - u_{N-1,j+1}^{-} \rb - 12 \kappa^2 \epsilon \lsq \lb u_{N+1,j+1}^{-} \rb^2 + \lb u_{N-1,j+1}^{-} \rb^2 - 2 u_{N+1,j+1}^{-} u_{N-1,j+1}^{-} \rsq \notag \\
&+ 8 h \epsilon \lb  u_{N+1,j+1}^{-} - u_{N-1,j+1}^{-} -  2 u_{N+1,j}^{-} + 2 u_{N-1,j}^{-} +  u_{N+1,j-1}^{-} - u_{N-1,j-1}^{-} \rb = \notag \\
&4 c^2 h \kappa^2 \lb u_{N+1,j+1}^{+} - u_{N-1,j+1}^{+} \rb - 12 \kappa^2 \epsilon \lsq \lb u_{N+1,j+1}^{+} \rb^2 + \lb u_{N-1,j+1}^{+} \rb^2 - 2 u_{N+1,j+1}^{+} u_{N-1,j+1}^{+} \rsq \notag \\
&+ 8 \frac{\beta}{c^2} h \epsilon \lb  u_{N+1,j+1}^{+} - u_{N-1,j+1}^{+} -  2 u_{N+1,j}^{+} + 2 u_{N-1,j}^{+} +  u_{N+1,j-1}^{+} - u_{N-1,j-1}^{+} \rb \label{fd_Con2}
\end{align}

As we are considering localised initial data for strains, if we take $L$ large enough then we can enforce zero boundary conditions for the strain i.e. $u_{x} = 0$. Therefore, applying a central difference approximation to this condition, we have
\begin{equation}
\frac{u_{1,j+1}^{-} - u_{-1,j+1}^{-}}{2h} = 0 \RA u_{1,j+1}^{-} = u_{-1,j+1}^{-} \eqtext{and} u_{2N+1,j+1}^{+} = u_{2N-1,j+1}^{+},
\label{fd_BC}
\end{equation}
and we now solve for the boundary points using these relations. We note that \eqref{fd_uminus} and \eqref{fd_uplus} form tridiagonal systems, specifically for the values $i=0,\dots,N$ and $i=N,\dots,2N$. Denoting the right-hand side of \eqref{fd_uminus} as $f_i$ and the right-hand side of \eqref{fd_uplus} as $g_i$, we make a rearrangement around the central boundary, namely
\begin{equation}
\tilde{f}_{N} = f_{N} + 2 \epsilon u_{N+1,j+1}^{-}, \quad \tilde{g}_{N} = g_{N} + 2 \epsilon \frac{\beta}{c^2} u_{N-1,j+1}^{+}.
\label{fgtilde}
\end{equation}
The equation system for $i=0,\dots,N$ can be written in matrix form as
 \begin{equation}
	\begin{pmatrix}
	 4 \epsilon + h^2 & - 4\epsilon && \\[0.3em]
	- 2\epsilon &  4 \epsilon + h^2 & - 2\epsilon & \\[0.3em]
	& \ddots &  \ddots  \\[0.3em]
	& - 2\epsilon &  4 \epsilon + h^2 & - 2\epsilon \\[0.3em]
	&& - 2\epsilon &  4 \epsilon + h^2
	\end{pmatrix}
	\begin{pmatrix}
	u_{0,j+1}^{-} \\[0.3em]
	u_{1,j+1}^{-} \\[0.3em]
	\vdots \\[0.3em]
	u_{N-1,j+1}^{-} \\[0.3em]
	u_{N,j+1}^{-}
	\end{pmatrix}
	=
	\begin{pmatrix}
	f_0 \\[0.3em]
	f_1 \\[0.3em]
	\vdots \\[0.3em]
	f_{N-1} \\[0.3em]
	\tilde{f}_{N}
	\end{pmatrix},
\label{fdSys}
\end{equation}
with a similar system for $i=N,\dots, 2N$. This system is tridiagonal and therefore we can use the Thomas algorithm (e.g. \cite{Ames77}) to solve both tridiagonal systems in terms of $u_{N+1,j+1}^{-}$ and $u_{N-1,j+1}^{+}$ respectively. This intermediary solution is then substituted into \eqref{fd_Con2} to obtain a nonlinear equation in terms of one of the ghost points, which can then be used to determine the solution for this time step. To obtain this expression, we denote
\begin{align}
u_{N,j+1}^{-} &= \phi_1^{-} + \psi_1^{-} u_{N+1,j+1}^{-}, &&u_{N,j+1}^{+} = \phi_1^{+} + \psi_1^{+} u_{N-1,j+1}^{+}, \notag \\
u_{N-1,j+1}^{-} &= \phi_2^{-} + \psi_2^{-} u_{N+1,j+1}^{-}, &&u_{N+1,j+1}^{+} = \phi_2^{+} + \psi_2^{+} u_{N-1,j+1}^{+},
\label{xat0}
\end{align}
where we calculate the coefficients $\phi_1^{\pm}$, $\psi_1^{\pm}$, $\phi_2^{\pm}$, $\psi_2^{\pm}$ from \eqref{fdSys} and the associated system for the second equation. To solve this system in terms of the ghost point $u_{N+1,j+1}^{-}$, we firstly express the variables $u_{N-1,j+1}^{+}$ and $u_{N+1,j+1}^{+}$ in terms of this ghost point. Therefore, denoting $u = u_{N+1,j+1}^{-}$ for brevity, we make use of \eqref{fd_Con1} and obtain
\begin{equation}
u_{N-1,j+1}^{+} = \frac{\phi_1^{-} - \phi_{1}^{+} + \psi_{1}^{-} u}{\psi_1^{+}} \eqtext{and} u_{N+1,j+1}^{+} = \phi_{2}^{+} + \frac{\psi_2^{+}}{\psi_1^{+}} \lb \phi_1^{-} - \phi_{1}^{+} + \psi_{1}^{-} u \rb.
\label{u_simp}
\end{equation}
Substituting \eqref{xat0} and \eqref{u_simp} into \eqref{fd_Con2} we can find a quadratic equation for $u$ of the form
\begin{equation}
h_0 \lb u_{N+1,j+1}^{-} \rb^2 + h_1 u_{N+1,j+1}^{-} + h_2 = 0,
\label{u_quad}
\end{equation}
where we have
\begin{align}
h_0 &= -12 \kappa^2 \epsilon \lsq \lb \psi_2^{-} \rb^2 - 2 \psi_2^{-} + 1 \rsq + 12 \kappa^2 \epsilon \lb \frac{\psi_1^{-}}{\psi_1^{+}} \rb^2 \lsq \lb \psi_2^{+} \rb^2 - 2 \psi_2^{+} + 1 \rsq, \notag \\
h_1 &=  \lb 1 - \psi_2^{-} \rb \lb 4 h \kappa^2 + 8 h \epsilon + 24 \kappa^2 \epsilon \phi_2^{-} \rb + \lb 1 - \psi_2^{+} \rb \lb \frac{\psi_1^{-}}{\psi_1^{+}} \rb \lb 4 c^2 h \kappa^2 + 8 \beta c^{-2} h \epsilon - 24 \kappa^2 \epsilon \phi_2^{+} \rb \notag \\
&\quad + 24 \kappa^2 \epsilon \frac{\psi_1^{-} \lb \phi_1^{-} - \phi_1^{+} \rb}{\lb \psi_1^{+}\rb^2}  \lsq \lb \psi_2^{+} \rb^2 - 2 \psi_2^{+} + 1 \rsq, \notag \\
h_2 &=  -\phi_{2}^{-} \lb 4 h \kappa^2  + 8 h \epsilon \rb - 12 \kappa^2 \epsilon \lb \phi_2^{-} \rb^2 + 8 h \epsilon \lb -  2 u_{N+1,j}^{-} + 2 u_{N-1,j}^{-} +  u_{N+1,j-1}^{-} - u_{N-1,j-1}^{-} \rb \notag \\
&\quad + 12 \kappa^2 \epsilon \lsq \lb \phi_2^{+} \rb^2 + \lb \frac{\phi_1^{-} - \phi_1^{+}}{\psi_1^{+}} \rb^2 \lsq \lb \psi_2^{+} \rb^2 - 2 \psi_2^{+} + 1 \rsq + 2 \lb \psi_2^{+} - 1 \rb \lb \frac{\phi_2^{+} \lb \phi_1^{-} - \phi_1^{+} \rb}{\psi_1^{+}}  \rb \rsq \notag \\
&\quad - \lb 4 c^2 h \kappa^2 + 8 \beta c^{-2} h \epsilon \rb \lsq \phi_2^{+} + \frac{\phi_1^{-} - \phi_1^{+}}{\psi_1^{+}} \lb \psi_2^{+} - 1 \rb \rsq  \notag \\
&\quad - 8 \beta c^{-2} h \epsilon \lb -  2 u_{N+1,j}^{+} + 2 u_{N-1,j}^{+} +  u_{N+1,j-1}^{+} - u_{N-1,j-1}^{+} \rb. \label{hcoef}
\end{align}
In order to simplify equations \eqref{hcoef}, we need to find expressions for the coefficients in \eqref{xat0}, so we consider the intermediary steps in the Thomas algorithm to obtain explicit representations. By considering the intermediary steps we can deduce that, for $N$ large enough, we can set $\psi_1^{-} = \psi_1^{+}$ and $\psi_2^{-} = \psi_2^{+}$ (this has been confirmed by numerical calculation) and therefore $h_0 \equiv 0$. This leads to the result
\begin{equation}
u_{N+1,j+1}^{-} = -\frac{h_2}{h_1},
\label{u_Nval}
\end{equation}
and therefore we can determine $u_{N,j+1}^{-}$, $u_{N,j+1}^{+}$ and similarly $u_{N-1,j+1}^{-}$ and $u_{N+1,j+1}^{+}$. These values can be substituted into the implicit solution of the tridiagonal system to determine the solution at each time step. It is worth noting that this scheme is second order.

To determine the initial condition, we differentiate \eqref{SB} and denote $\varepsilon (x,t) = u_{x} (x,t)$. The initial condition is then taken as the exact solitary wave solution of this equation for $\varepsilon (x,t)$, and therefore the exact ``kink'' solution for \eqref{SB}. Explicitly we have
\begin{align}
u \lb x, 0 \rb^{\pm} &= - \frac{v_1 \sqrt{v_1^2 - 1}}{\sqrt{2 \epsilon}} \lsq \tanh{\lb \frac{\sqrt{v_1^2 - 1}}{2 v_1 \sqrt{2 \epsilon}} x \rb} - 1 \rsq, \notag \\
u \lb x, \kappa \rb^{\pm} &= - \frac{v_1 \sqrt{v_1^2 - 1}}{\sqrt{2 \epsilon}} \lsq \tanh{ \lb \frac{\sqrt{v_1^2 - 1}}{2 v_1 \sqrt{2 \epsilon}} \lb x - v_1 \kappa \rb \rb} - 1 \rsq,
\label{BSIC}
\end{align}
where $v_1$ is the velocity of the wave. If the initial condition was not given by an explicit analytic function then we could deduce the second initial condition for the scheme by taking the forward difference approximation (simulations have shown that either case is sufficiently accurate).

\subsection{Semi-Analytical Approach}
\label{sec:KNum}
To solve equations \eqref{IKdV}, \eqref{RKdV} and \eqref{TKdV} numerically, we use the SSPRK(5,4) scheme as described in \cite{Gottlieb09} (see \cite{Obregon14} for example of this method). A hybrid Runge-Kutta scheme was used initially (e.g. \cite{Marchant02} and the methods used for the finite-difference terms in the equation were based upon \cite{Chu83, Vliegenthart71}). However it was found the SSPRK(5,4) scheme had a higher accuracy, due to its stability preserving properties, and the computational time was similar for both schemes. In addition, the SSPRK(5,4) scheme can be adapted more easily.

We discretise the domain $\lsq -L, L \rsq \times \lsq 0, T \rsq$ into a grid with equal spacings $h = \Delta \xi$, $\kappa = \Delta X$, and the analytical solution $e^{\pm} \lb x, t \rb$ is approximated by the exact solution of the numerical scheme, $e^{\pm} \lb ih, j \kappa \rb$ (denoted $e_{i,j}^{\pm}$), where $e_{i,j}^{-}$ is the sum of the numerical solutions to \eqref{IKdV} and \eqref{RKdV} and $e_{i,j}^{+}$ is the numerical solution to \eqref{TKdV}.

The SSPRK(5,4) scheme is as follows. Let $e(x,t)$ denote a solution of the KdV equation. Given that the solution at time $t_{j} = j\kappa$ is given by
\begin{equation}
e_{i,j} = e \lb ih, j \kappa \rb, \quad i = 0, 1, \dots, 2N,
\end{equation}
then the solution at $t_{j+1} = \lb j  + 1 \rb \kappa$ is given by
\begin{align*}
e^{(1)} &= e_{i,j} + 0.391752226571890 \kappa F \lb e_{i,j} \rb, \\
e^{(2)} &= 0.444370493651235 e_{i,j} + 0.555629506348765 e^{(1)} + 0.368410593050371 \kappa F \lb e^{(1)} \rb, \\
e^{(3)} &= 0.620101851488403 e_{i,j} + 0.379898148511597 e^{(2)} + 0.251891774271694 \kappa F \lb e^{(2)} \rb, \\
e^{(4)} &= 0.178079954393132 e_{i,j} + 0.821920045606868 e^{(3)} + 0.544974750228521 \kappa F \lb e^{(3)} \rb, \\
e_{i,j+1} &= 0.517231671970585 e^{(2)} + 0.096059710526147 e^{(3)} + 0.063692468666290 \kappa F \lb e^{(3)} \rb \\
&\quad + 0.386708617503269 e^{(4)} + 0.226007483236906 \kappa F \lb e^{(4)} \rb,
\end{align*}
where the function $F$ is the finite-differenced form of all the terms in the KdV equation involving spatial derivatives. Note that the coefficients here are chosen in such a way to optimise the time step at each point and, due to the complexity of each coefficient, are presented to 15 decimal places.

To obtain an expression for $F$, central difference approximations are applied for the first and third derivatives and an average is taken for the nonlinear term (as was performed by Zabusky and Kruskal, see \cite{Zabusky65, Drazin89}). We also assume boundary conditions of the form
\begin{equation*}
e(\pm L,t) = 0, \quad e_{x}( \pm L,t) = 0 \RA e_{0,j} = e_{1,j} = e_{2N-1,j} = e_{2N,j} = 0.
\end{equation*}
Using this scheme in equation \eqref{IKdV} (which is identical to the equation \eqref{RKdV} in characteristic variables) we obtain
\begin{align}
F \lb p_{i,j} \rb &= \lb p_{i+1,j} + p_{i,j} + p_{i-1,j} \rb  \lb p_{i+1,j} - p_{i-1,j} \rb \notag \\
& \quad - \frac{1}{2 h^3} \lb p_{i+2,j} - 2p_{i+1,j} + 2p_{i-1,j} - p_{i-2,j} \rb, \quad i = 2, 3, \dots, 2N-2,\label{fd_IKdV}
\end{align}
and for \eqref{TKdV} we obtain
\begin{align}
F \lb p_{i,j} \rb &= \frac{1}{c^2} \lb p_{i+1,j} + p_{i,j} + p_{i-1,j} \rb  \lb p_{i+1,j} - p_{i-1,j} \rb \notag \\
& \quad - \frac{\beta}{2 c^2 h^3} \lb p_{i+2,j} - 2p_{i+1,j} + 2p_{i-1,j} - p_{i-2,j} \rb, \quad i = 2, 3, \dots, 2N-2. \label{fd_TKdV}
\end{align}
The initial condition is taken as the exact solution of the KdV equation \eqref{IKdV}, namely
\begin{equation}
I \lb x, 0 \rb = -\frac{v}{2} \sechn{2}{\frac{\sqrt{v}}{2} x },
\label{KdVIC}
\end{equation}
where $v$ is the velocity of the soliton and can be related to the velocity of the solution \eqref{BSIC} by the formula $v_1 = \sqrt{1 + 2 \epsilon v}$. For equation \eqref{TKdV} a multiplicative factor is applied to the initial condition in \eqref{KdVIC}, namely that which we derived in \eqref{CT}. Explicitly we have
\begin{equation}
T \lb x, 0 \rb = -\frac{v}{c \lb c + 1 \rb} \sechn{2}{\frac{\sqrt{v}}{2}  x }.
\label{TKdVIC}
\end{equation}
Similarly, for equation \eqref{RKdV} a different multiplicative factor is applied to the initial condition \eqref{KdVIC} of the form \eqref{CR} and therefore we have
\begin{equation}
R \lb x, 0 \rb = -\frac{v \lb c - 1 \rb}{2 \lb c + 1 \rb} \sechn{2}{\frac{\sqrt{v}}{2} x }.
\label{RKdVIC}
\end{equation}
 
\section{Results}
\label{sec:Res}
In what follows, we let $\epsilon = 0.01, v = 0.35$ and  take a step size of $h = \kappa = 0.01$ in the finite-difference scheme and a step size of $h = 0.1, \kappa = 0.001$ in the SSPRK(5,4) scheme. The value of $h$ and $\kappa$ chosen for the SSPRK(5,4) scheme was chosen as the same step size taken for the hybrid Runge-Kutta scheme in \cite{Marchant02}. Subsequent numerical experiments have shown that the SSPRK(5,4) scheme is not sensitive to considerable changes in the step sizes, and the results for these chosen step sizes and much larger values of $h = 0.25$, $\kappa = 0.01$ produce comparable results with an error of $6.5 \times 10^{-3}$. As shown in \cite{Khusnutdinova11}, the finite-difference method for the Boussinesq equation is  linearly stable (using a von Neumann linear stability analysis) for values of $\kappa$ satisfying
\begin{equation}
\kappa < \kappa_c = \sqrt{\frac{h^2 + 4\beta c^{-2}}{c^2 + f_0}},
\end{equation}
where $f_0$ is the constant used in the linearised scheme. In practice the stricter condition of
\begin{equation}
\kappa  < \frac{1}{2} \kappa_c,
\end{equation}
is imposed, to help accommodate for the nonlinearity effects. The values of $h$ and $\kappa$ chosen above satisfy this relation.

The results are plotted for the strain waves, to correspond with the prescription of the initial condition. Therefore, we denote $e^{-} = u_{x}^{-}$ and $e^{+} = u_{x}^{+}$.

\subsection{Test Cases}
\begin{figure}
\centering
\includegraphics[scale=0.34, trim = 0mm 30mm 0mm 55mm]{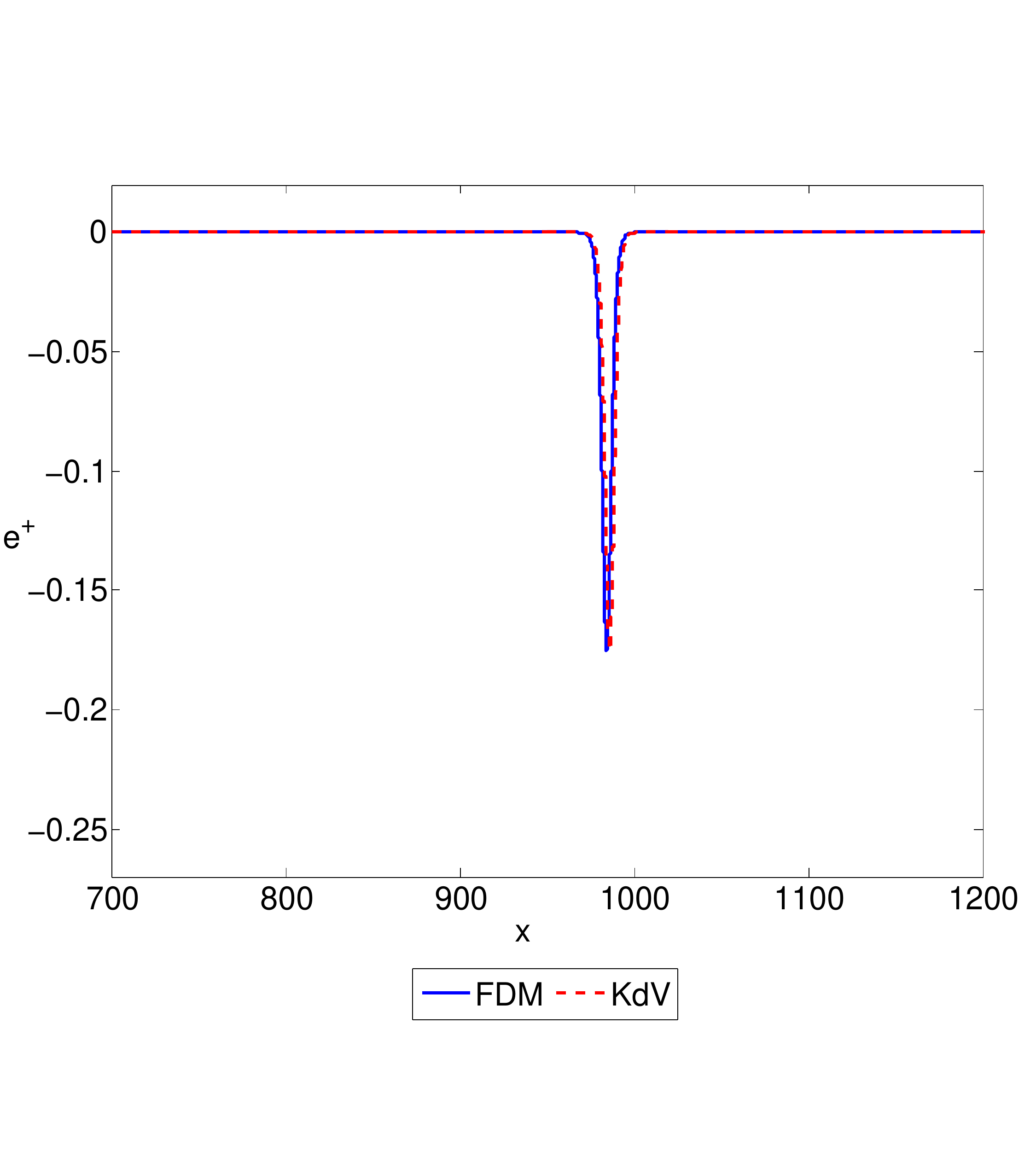}
\caption{The solution $e^{+}$ at $t=1000$ in the equations \eqref{fd_uplus} and \eqref{fd_TKdV} with exact initial conditions and initial position $x=-50$.}
\label{SST}
\end{figure}

\begin{figure}
\centering
\includegraphics[scale=0.34, trim = 0mm 25mm 0mm 45mm]{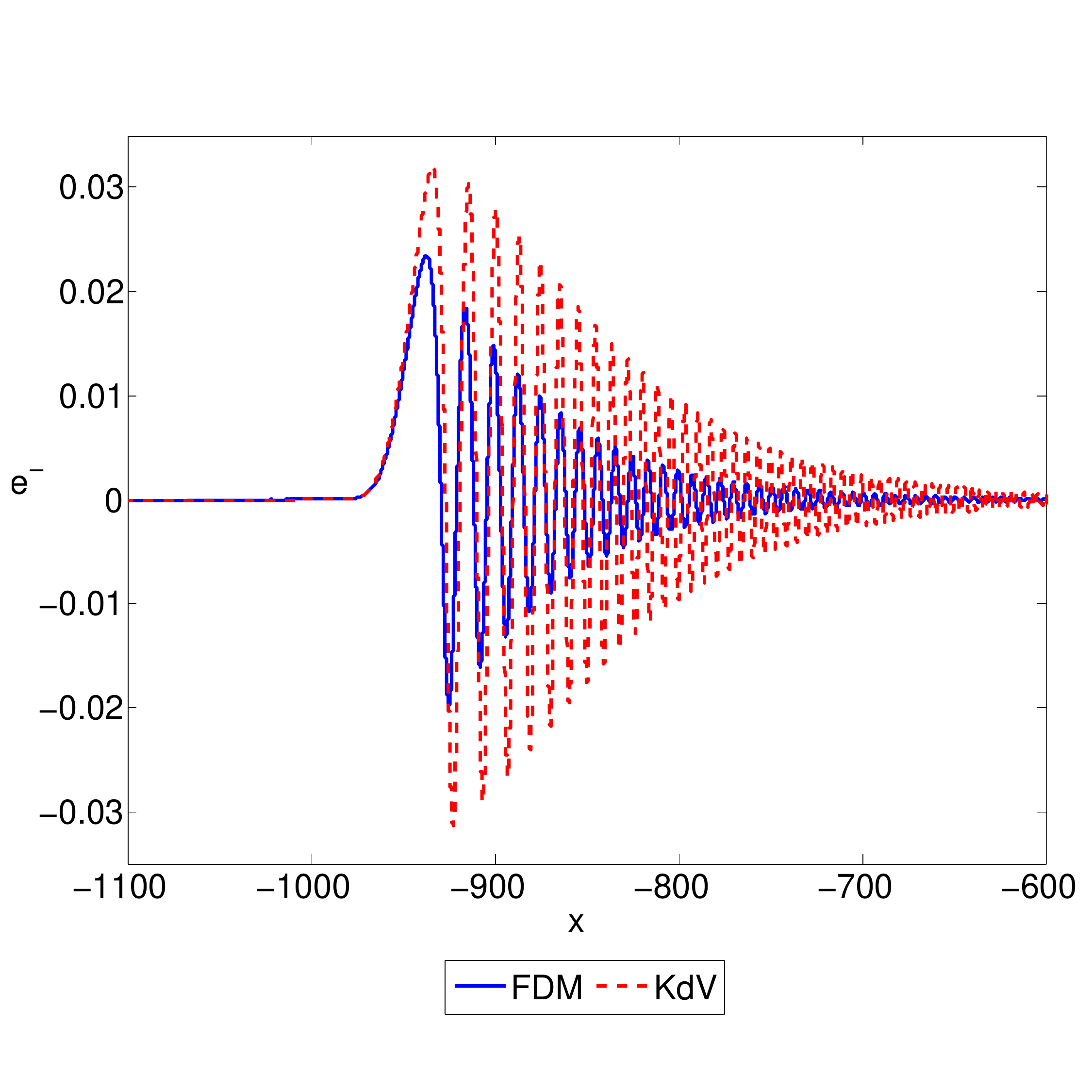}
\caption{The solution $e^{-}$ at $t=1000$ in the equations \eqref{fd_uminus} and \eqref{fd_IKdV} with $c=0.25$ and an initial position of $x=-50$.}
\label{LowCR}
\end{figure}

Firstly we consider the case where $\beta = c = 1$. The initial condition for the Boussinesq equation \eqref{SB} is given by \eqref{BSIC}. We solve the equation for the displacement $u^+(x,t)$ and  plot the strain $u^+_x (x, t)$. The initial condition for the KdV equation \eqref{TKdV} is provided by \eqref{TKdVIC}.  Viewing the solution of the first scheme as exact, the solution of the second scheme will be accurate to leading order, with a small correction to the wave at $O \lb \epsilon^2 \rb$. The solution for $e^{-}$ at the initial time and for $e^{+}$ at a sufficiently large time should describe the same right-propagating solitary wave. The result of the calculation for this case, for both numerical schemes, is presented in Figure \ref{SST} for the transmitted wave. As $c=1$, there is no leading order reflected wave. Figure \ref{SST} shows a good agreement between the solutions, with a small phase shift between the two cases. This can be remedied by adding higher order terms to the solution of the second scheme.
Another test case is for a value of $c$ such that $c<1$. Considering the expressions for $C_R$ and $C_T$, that is
\begin{equation*}
C_R = \frac{c-1}{c+1}, \quad C_T = \frac{2}{c \lb 1 + c \rb},
\end{equation*}
and noting that $c>0$ (a physical requirement), we observe that $C_T > 0$ for all $c$, while $C_R < 0$ for $c<1$ and $C_R > 0$ for $c>1$. Recalling our observations from Section \ref{sec:WNL} we expect that, for a value of $c<1$, a dispersive wave train will be present in the reflected wave field. For $c=0.25$, we note that $C_R = -3/5$ and the result for this calculation is presented in Figure \ref{LowCR}. It can be seen that a dispersive wave train is present in both numerical schemes. The KdV approximation overestimates the amplitude of the wave train, but correctly resolves all its main features. It is worth noting that this solution looks like the Airy function solution of the linearised KdV equation, which is the small amplitude limit of the similarity solution of the KdV equation related to the Painleve II equation (see \cite{Ablowitz11, Ablowitz81} and references therein for more details).

\begin{figure}
\centering
\includegraphics[scale=0.34, trim = 0mm 25mm 0mm 35mm]{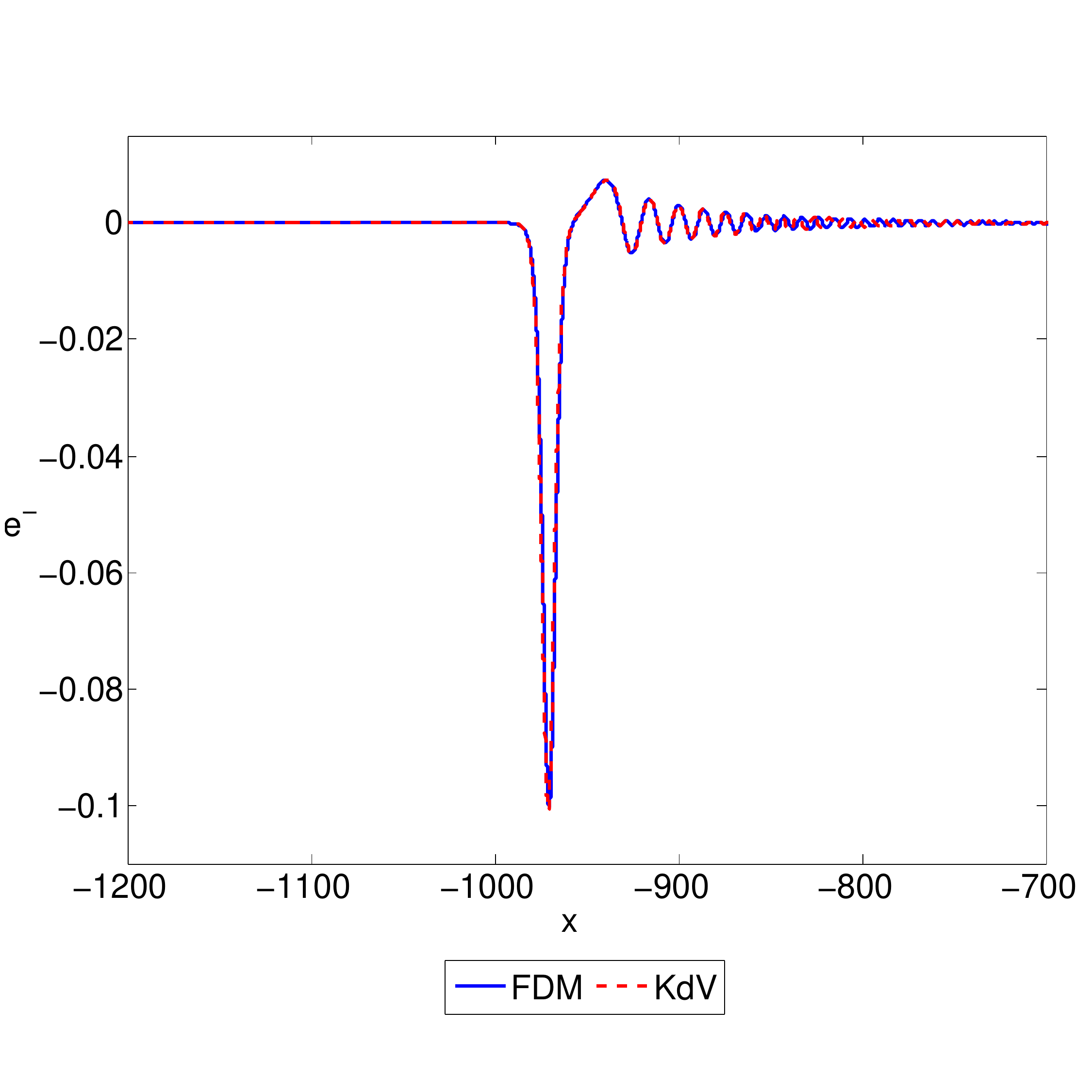}
\caption{The solution $e^{-}$ at $t=1000$ in the equations \eqref{fd_uminus} and \eqref{fd_IKdV} with $c=5$ and initial position $x=-50$.
\label{HighCR}
}
\includegraphics[scale=0.34, trim = 0mm 45mm 0mm 35mm]{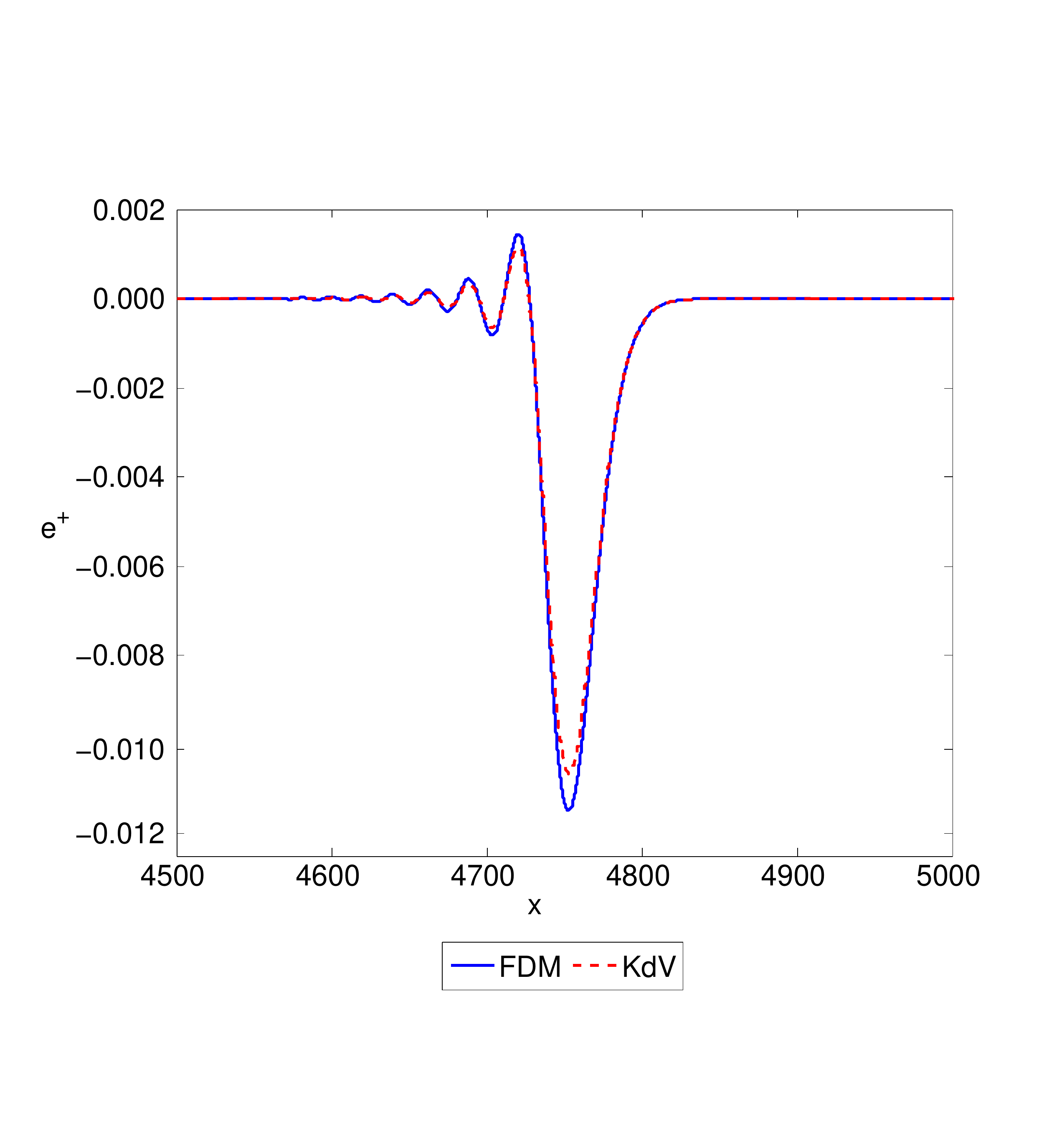}
\caption{The solution $e^{+}$ at $t=1000$ in the equations \eqref{fd_uplus} and \eqref{fd_TKdV} with $c=5$ and initial position $x=-50$.
\label{HighCT}
}
\end{figure}

The next case to consider is for large values of $c$. Using the results of Section \ref{sec:WNL}, specifically \eqref{CR} and \eqref{CT}, we would expect the leading order reflected wave to be closer in amplitude to the initial wave and the transmitted wave to be of a much smaller amplitude. An example is presented in Figure \ref{HighCR} and Figure \ref{HighCT} for $c=5$, where we have
\begin{equation*}
C_{R} = \frac{2}{3}, \quad C_{T} = \frac{1}{15},
\end{equation*}
and we can see from the coefficients that we would expect the reflected wave to have a larger amplitude than the transmitted wave, by approximately one order of magnitude in this case.

\subsection{Predictions}
\begin{figure}
\centering
\includegraphics[scale=0.34, trim = 0mm 20mm 0mm 45mm]{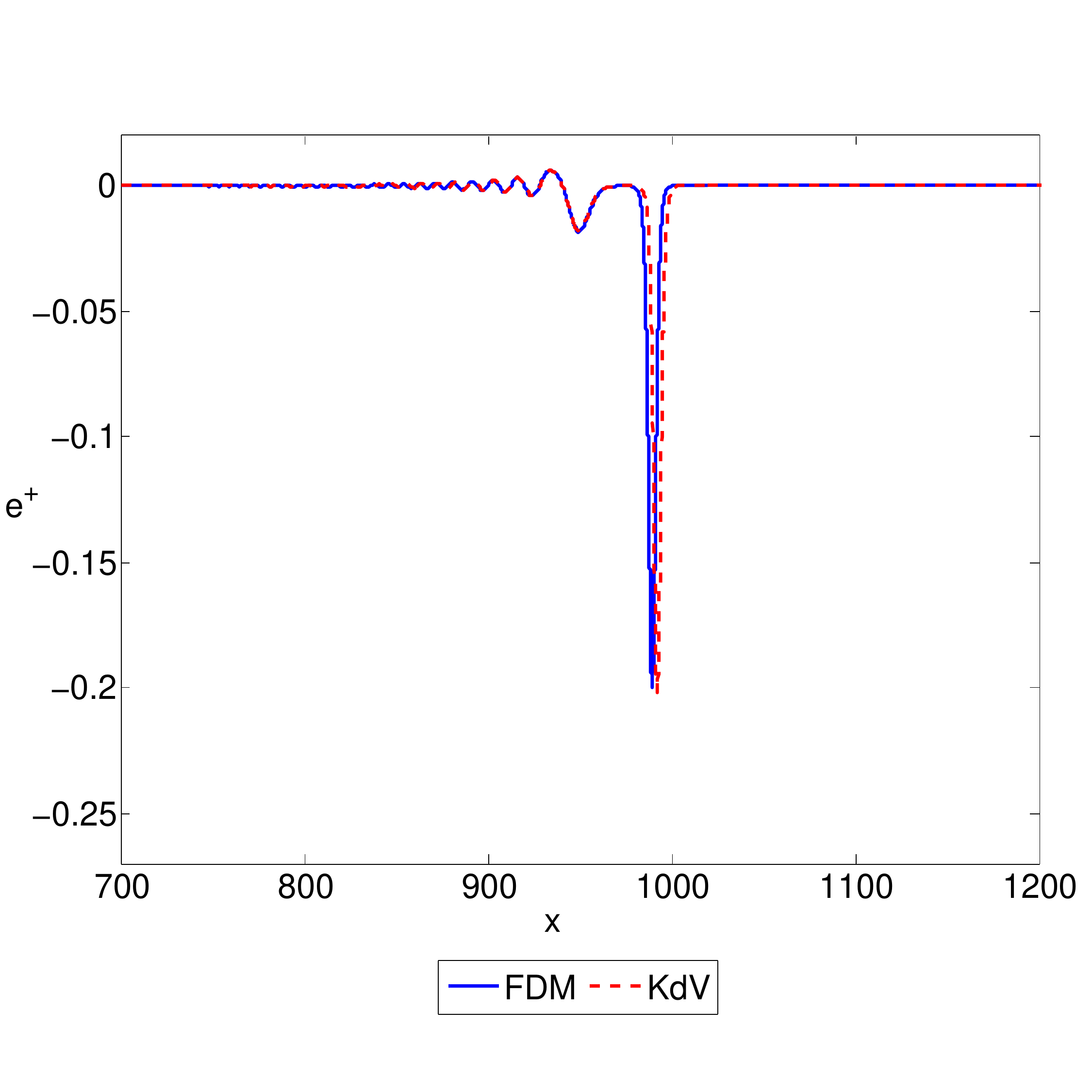}
\caption{The solution $e^{+}$ at $t=1000$ in the equations \eqref{fd_uplus} and \eqref{fd_TKdV} with $\beta=5/8$, corresponding to $n=2$ and $k=1$, and initial position $x=-50$.
\label{n2k1T}
}
\includegraphics[scale=0.34, trim = 0mm 20mm 0mm 15mm]{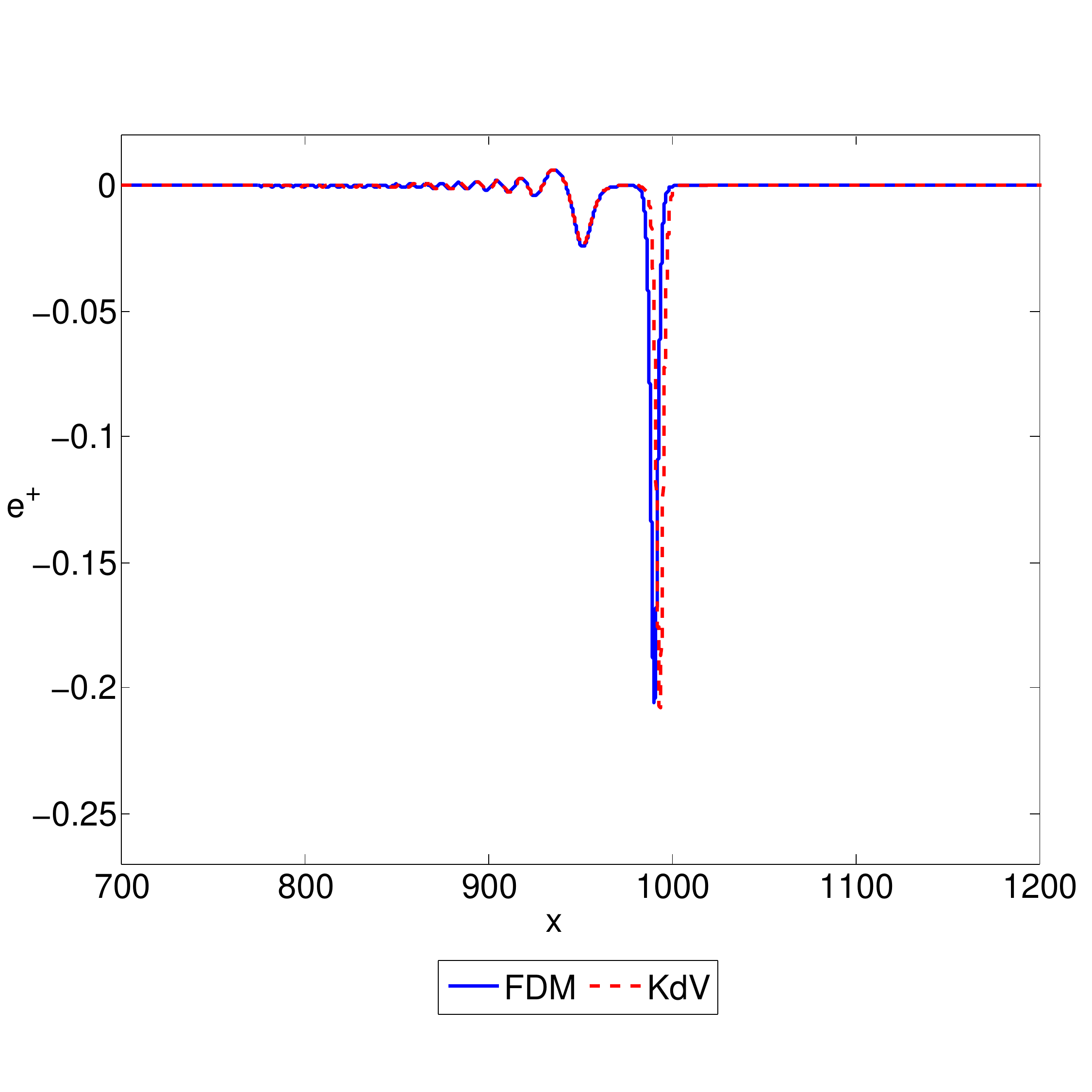}
\caption{The solution $e^{+}$ at $t=1000$ in the equations \eqref{fd_uplus} and \eqref{fd_TKdV} with $\beta=5/9$, corresponding to $n=3$ and $k=1$, and initial position $x=-50$.
\label{n3k1T}
}
\end{figure}

\begin{figure}
\centering
\includegraphics[scale=0.34, trim = 0mm 20mm 0mm 45mm]{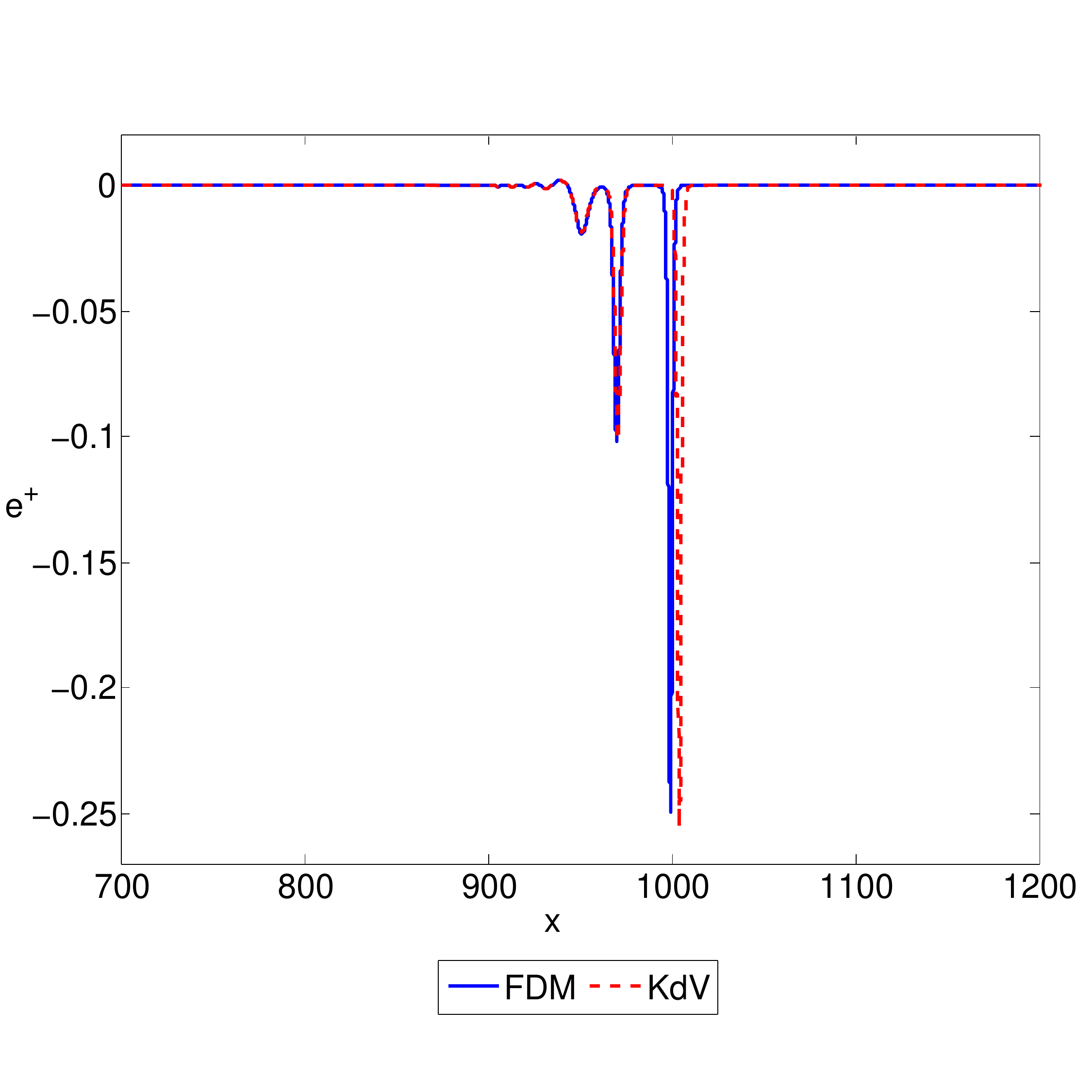}
\caption{The solution $e^{+}$ at $t=1000$ in the equations \eqref{fd_uplus} and \eqref{fd_TKdV} with $\beta=1/5$, corresponding to $n=3$ and $k=3$, and initial position $x=-50$.
\label{n3k3T}
}
\includegraphics[scale=0.34, trim = 0mm 20mm 0mm 15mm]{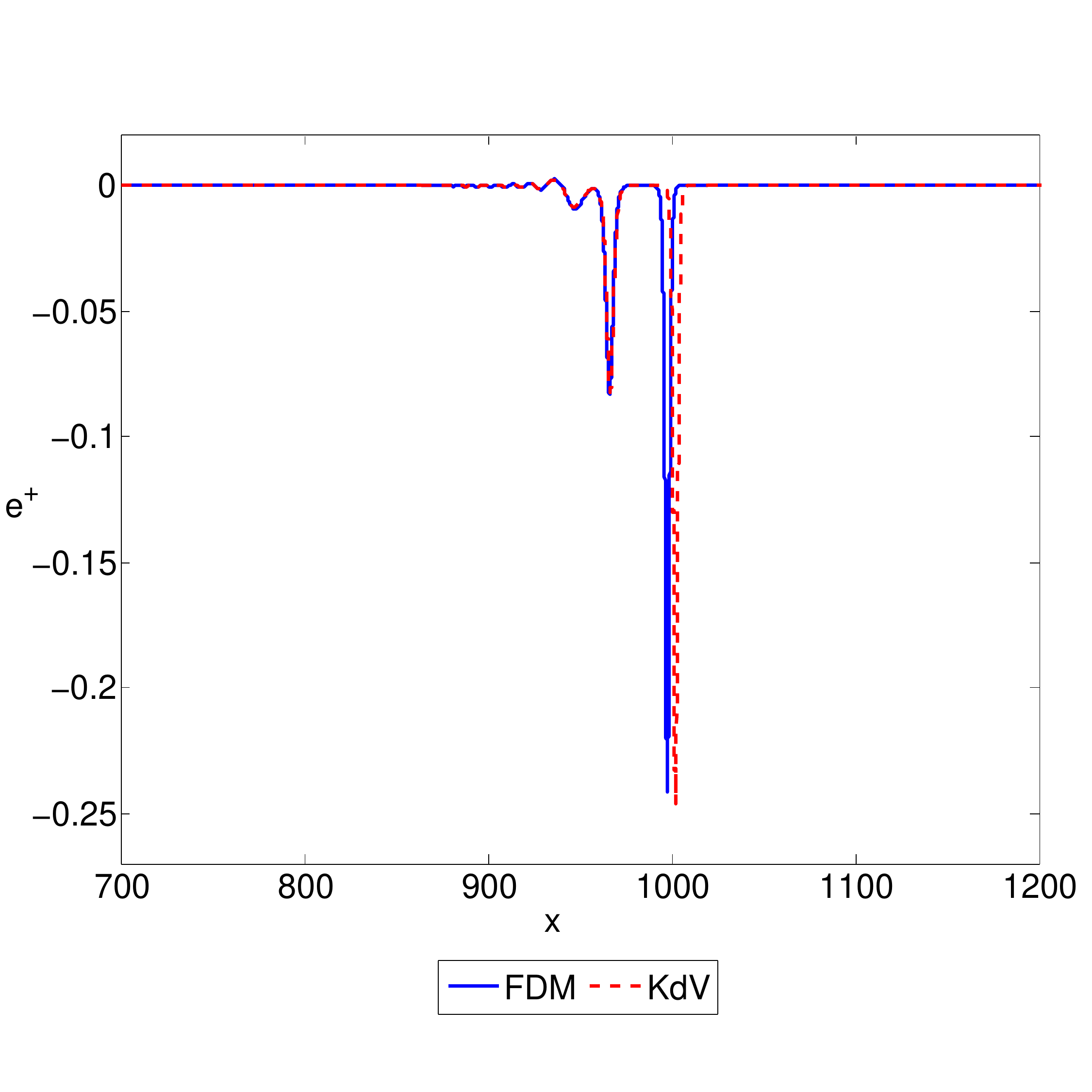}
\caption{The solution $e^{+}$ at $t=1000$ in the equations \eqref{fd_uplus} and \eqref{fd_TKdV} with $\beta=1/4$, corresponding to $n=4$ and $k=2$, and initial position $x=-50$.
\label{n4k2T}
}
\end{figure}

Following the scheme outlined in Section \ref{sec:Fis}, we define $c = 1$ and recall that
\begin{equation}
\beta = \beta(n,k) = \frac{n^2 + k^2}{n^2 \lb 1 + k^2 \rb}.
\end{equation}
We now consider the behaviour in our problem formulation using these parameters. The initial condition remains the same and we change the value of $n$ and $k$ to obtain different cases. Table 1 in \cite{Khusnutdinova08} presents results for 9 different configurations and all of these configurations have been checked. For brevity, we only present results for 4 cases. The expected number of solitary waves in the transmitted wave field in these cases are summarised in Table \ref{tab:sol}. The results are presented in Figures \ref{n2k1T} - \ref{n4k2T}. In each of the following cases we note that there is no leading order reflected wave, as $c=1$, and therefore only the transmitted wave field is presented. We observe that the number of predicted solitons is in agreement with the numerics in all cases. In the cases where the smaller solitons are close in amplitude to the radiation, a further simulation was run for a larger time to ensure that the smaller soliton moves away from the radiation generated by the continuous spectrum. The figures are all presented at the same time value of $t=1000$ for comparison.

\begin{table}
\vspace{10pt}
\centering
\begin{tabularx}{0.85\textwidth}{| Y | Y | Y | Y |}
\hline
Number of layers, $n$ & Ratio of height and half width, $k$ & Value of dispersion coefficient, $\beta$ &Number of solitons \\ \hhline{|=|=|=|=|}
2 & 1 & 5/8 & 2 \\ \hline
3 & 1 & 5/9 & 2 \\ \hline
3 & 3 & 1/5 & 3 \\ \hline
4 & 2 & 1/4 & 3 \\ \hline
\end{tabularx}
\caption{Predictions on the number of solitons present in the transmitted wave field in the delaminated section of the bar, for various choices of $n$ and $k$.}
\label{tab:sol}
\vspace{2pt}
\end{table}

\subsection{Predicted Heights}
A prediction for the amplitude of the lead soliton is provided in \cite{Khusnutdinova08}, namely that
the ratio of its amplitude to that of the incident soliton is
\begin{equation}
C_a = \frac{\beta}{4} \lb \sqrt{1 + \frac{8}{\beta}} - 1 \rb^2.
\end{equation}
We consider the combination of values $\kappa = 1, 2, 3$ and $n = 2, 3, 4$ for the following results. Taking $h = k = 0.01$  in the finite-difference scheme we have a maximum error of $5.877 \times 10^{-3}$ and for step sizes $h = 0.1, k = 0.001$ in the SSPRK(5,4) scheme we have a maximum error of $1.461 \times 10^{-4}$.

\subsection{Comparison of Schemes}
Reviewing the results presented here, it can be seen that the semi-analytical approach produces results comparable to the direct finite-difference scheme for many cases, particularly for long waves. As the model is a long-wave model, this is the desirable behaviour for the schemes. Crucially, the semi-analytical approach requires the solving of, at most, two equations in each section of a bar (reflected and transmitted) while the direct finite-difference method requires the solution of multiple tridiagonal equations systems, and the solution of the nonlinear equation at $x=0$ has to be substituted into the implicit solution at all other points. Therefore, the semi-analytical approach is more desirable as additional sections are included in the bar. In addition, an analysis of the time required to calculate a solution by each scheme shows that the semi-analytical approach is faster.\footnote{Both methods were programmed in C and compiled using the Intel compiler. The spatial domain for the semi-analytical method used 75,000 points, while the spatial domain for the finite-difference method used 400,000 points. When run on a 2.66 GHz Intel Core i5 machine, the approximate CPU time for the direct finite-difference scheme was 12 hours, compared to 4 hours for the semi-analytical method. Each scheme was parallelised and therefore the actual calculation time was reduced by a factor based upon the number of cores in the processor.

Subsequent numerical experiments have shown that in the simple settings considered in the present paper the domain size can be further reduced to improve the calculation time, so we can consider a spatial domain containing 15,000 points for the semi-analytical method, and 150,000 points for the finite-difference method. This reduced the calculation time to approximately 20 minutes for the finite-difference scheme and 15 minutes for the semi-analytical scheme. However, as the semi-analytical scheme is not sensitive to considerable increases in the step sizes, we can take step sizes of $h=0.25$ and $\kappa=0.01$ to obtain a calculation time of 34 seconds. Therefore the resulting calculation in the semi-analytical scheme was 35 times faster than the finite-difference scheme.} The accuracy of this method can be improved further by adding higher-order terms \cite{Khusnutdinova12, Khusnutdinova14}. It must be noted that, in these papers, we have developed a semi-analytical numerical approach to the solution of the initial value problem for Boussinesq-type equations, however the advantages of the semi-analytical method compared to the standard methods were not obvious. They became apparent for the type of problems discussed in this paper.

\section{Conclusions}
\label{sec:Conc}
In this paper we considered two numerical schemes for solving two boundary-value problems matched at $x=0$. This problem represents the propagation of a strain wave in a bar with a perfect bonding between the two layers for the first boundary-value problem, and a bar with complete delamination in the second boundary-value problem.

At this point we looked for a weakly nonlinear solution by taking a multiple-scales expansion in terms of the appropriate set of fast and slow variables. This produced three KdV equations satisfying initial-value problems, describing the leading order incident, reflected and transmitted waves. The initial values in these equations are fully described in terms of the leading order incident wave, with reflection and transmission coefficients being derived to describe these initial conditions. It was noted that a bar made of one and the same material will not generate a leading order reflected wave. In addition, expressions were found for the higher order corrections in terms of the leading order incident and reflected waves, and the reflection and transmission coefficients.

Following \cite{Khusnutdinova08}, the process of fission of an incident solitary wave was then discussed and expressions for the number of solitons in the delaminated section and their eigenvalues were found. This was  applied to the case when the materials in the perfectly bonded and delaminated sections of the bar are one and the same, and a formula for the number of solitons present in the delaminated section was found in terms of the geometry of the waveguide. This allowed for predictions to be made to the number of solitons for a given configuration of the bar. A  similar description was given for the reflected wave and a similar prediction can be made in this case.

Two numerical schemes were suggested to describe this behaviour, taking account of the original derivation and the weakly nonlinear approach. For the original equations we made use of finite-difference techniques, which resulted in two tridiagonal systems and a nonlinear difference equation linking the systems. These systems were solved,  in terms of ``ghost points'' in both systems, using a Thomas algorithm \cite{Ames77} and the result of this calculation was used to find the solution of the nonlinear equation for one of the ghost points. The solution for this ghost point is then substituted back into the implicit solution of the tridiagonal system to determine the solution at a given time value. The result allowed us to analyse the leading order transmitted and reflected waves in their respective domains and compare the results to known analytical predictions.  The numerical modelling has confirmed that the incident soliton fissions in the delaminated area, which can be used to detect the defect.

The second scheme solves the derived KdV equations in terms of the incident strain wave. This semi-analytical approach makes use of a SSPRK(5,4) scheme, with the finite-differenced form of the spatial derivatives used as the function in the scheme \cite{Gottlieb09}. This scheme was used to calculate the incident, transmitted and reflected waves from their respective equations. The reflection and transmission coefficients determine the magnitude of the initial condition in the equations for the reflected and transmitted waves respectively.

These results were then compared to each other and to known analytical results. It was found that the schemes were in agreement to leading order, with a slight phase shift between the two solutions representing the $O \lb \epsilon^2 \rb$ difference between their propagation speeds. Furthermore, the numerics were in agreement with the predicted theory for the number of solitons in the delaminated section for several choices of waveguide geometry. In addition, it was found that the semi-analytical approach has a smaller computation time than the direct method, and is simpler to implement in comparison to the direct finite-difference approach. This effect will be more dramatic as the number of equations is increased (corresponding to more sections in the bar).
Finally, the methods used in this paper can be generalised to a bar with a soft bonding layer, which leads to a system of coupled Boussinesq equations for $x<0$ and uncoupled equations for $x>0$ \cite{Khusnutdinova09}.
\bigskip

\noindent {\color{blue}Acknowledgements.} 
\bigskip

The authors would like to thank Noel Smyth and the referees for useful discussions of the finite-difference approach. MRT is supported by an EPSRC studentship.

\bibliographystyle{vancouver}
\bibliography{Research}

\end{document}